\documentstyle[aps,prbbib,twocolumn,epsf]{revtex}

\begin{document}

\draft
\title{Gate-controllable spin-battery} 

\author{Wen Long}

\address{
$^1$Department of Physics, Capital Normal University, Beijing 100037, 
China}

\author{Qing-feng Sun and Hong Guo }

\address{
Center for the Physics of Materials and Department
of Physics, McGill University, Montreal, PQ, Canada H3A 2T8.\\
International Center for Quantum Structures, Institute of Physics,
Chinese Academy of Sciences, Beijing 100080, China }

\author{Jian Wang}

\address{
Department of Physics, The University of Hong Kong, Pokfulam Rood,
Hong Kong, China}

\maketitle

\begin{abstract}

We propose a gate-controllable spin-battery for spin current. The spin-battery
consists of a lateral double quantum dot under a uniform magnetic field. A 
finite DC spin-current is driven out of the device by controlling a set  of 
gate voltages. Spin-current can also be delivered in the absence of 
charge-current. The proposed device should be realizable using present 
technology at low temperature.

\end{abstract}

To be able to generate and control spin-current is of great importance for
spintronics\cite{ref1}. Traditionally, spin injection from a ferromagnetic 
material to a normal metal or semiconductor material has been used to obtain 
spin polarized charge-current. Spin injection into non-Fermi 
liquid\cite{addref1} as well as by circularly polarized light\cite{addref2}
have also been investigated. More recently, several theoretical proposals 
for spin-battery were reported for the generation of pure spin-current without 
charge-current.\cite{ref2,ref3,ref4} The idea is that when spin-up electrons 
move to one direction while an equal number of spin-down electrons move to 
the opposite direction, the net charge-current 
$I_e =e(I_{\uparrow} +I_{\downarrow})$ vanishes and a finite spin-current
$I_s =\frac{\hbar}{2}(I_{\uparrow} -I_{\downarrow})$ emerges. Here
$I_{\uparrow}$ ($I_{\downarrow}$) is the spin-up (spin-down) electron current.
Although conceptually interesting, existing spin-battery proposals all
involve time dependent external fields\cite{ref2,ref3,ref4} which make
practical realization somewhat complicated. It is the purpose of this paper 
to propose and investigate a novel spin-battery design which is gate 
controllable involving no time varying fields.

The gate controllable spin-battery is schematically shown in Fig.(1). It 
consists of a lateral double quantum-dot fabricated in two-dimensional 
electron gas (2DEG) with split gate technology. The two QDs are coupled to 
three leads: lead-1 and 3 couple to one QD each, lead-2 couples to both.
The two QDs are separated by a high potential barrier so that 
tunnel coupling between them can be neglected. To distinguish spin of the 
electrons, a magnetic field $B$ is applied to the QDs
to induce a Zeeman splitting. Two gate voltages $V_{g,\alpha}$ control 
energy levels of the $\alpha$-th QD, where $\alpha =upper,lower$ $(u,l)$,
indicating the upper and lower QD of Fig.(1). Finally, the terminal voltages 
for the three leads are set such that $V_1 >V_2 >V_3$ (Fig.2), they
provide energy source for the spin-battery.

Before presenting results, we first discuss why the system of Fig.(1) 
can deliver a spin-current. Due to field $B$, a spin degenerate 
level $\epsilon_{\alpha}$ of the $\alpha$-QD is split into spin-up/down
levels $\epsilon_{\alpha\uparrow}/\epsilon_{\alpha\downarrow}$. Let's assume 
$\epsilon_{\alpha\uparrow} < \epsilon_{\alpha\downarrow}$.
By adjusting gate voltages $V_{g,\alpha}$, we shift these levels. 
In particular, we set $V_{g,lower}$ such that electron occupation number 
in the lower-QD is changing between 0 and 1 (even to odd), with the level 
$\epsilon_{lower,\uparrow}$ locating between $\mu_1$ and $\mu_2$,
where $\mu_i =eV_i$ is the chemical potential of lead $i$. Similarly, 
we set $V_{g,upper}$ such that the upper-QD has an electron occupying 
state $\epsilon_{upper,\uparrow}$, while the other state 
$\epsilon_{upper,\downarrow}$ is pushed to higher energy 
$\epsilon_{upper,\downarrow} +U$ due to Coulomb interaction $U$.  This way, 
the electron occupation number in the upper-QD is changing between 
1 and 2 (odd to even), and the level $\epsilon_{upper,\downarrow} +U$ locates 
between $\mu_2$ and $\mu_3$. The energy level diagram shown in Fig.(2) 
is now established. From Fig.(2), it is clear that a spin-up electron 
in lead-1 can tunnel into the lower-QD and further to lead-2. Similarly, 
a spin-down electron in lead-2 can tunnel into the upper-QD and flows 
to lead-3. Therefore, in lead-2 spin-up electrons flow in and 
spin-down electrons flow out: they move in opposite directions so that 
a net spin-current is generated. Hence, by adjusting gate potentials 
the device of Fig.(1) generates a spin-current in the region labeled by (A,B).

We now present detailed analysis. The lateral double-QD device is 
described by the following Hamiltonian:
\begin{eqnarray}
& H & = 
\sum\limits_{\alpha\sigma}
\left(\epsilon_{\alpha} -\sigma g\mu B_{\alpha}/2 \right)
d^{\dagger}_{\alpha\sigma} d_{\alpha\sigma}  
+ \sum\limits_{\alpha} U_{\alpha} 
d^{\dagger}_{\alpha \uparrow}  d_{\alpha \uparrow}
d^{\dagger}_{\alpha \downarrow}  d_{\alpha \downarrow} \nonumber \\
& &
+ \sum\limits_{nk\sigma}
 \epsilon_{nk} a^{\dagger}_{n k\sigma} a_{n k\sigma} 
 +  \sum\limits_{nk\sigma\alpha} \left( t_{n,\alpha } 
a^{\dagger}_{n k\sigma} d_{\alpha\sigma} 
+ H.c. \right)  
\label{hamiltonian}
\end{eqnarray}
where $a^{\dagger}_{nk\sigma}$($a_{n k\sigma}$) and
$d^{\dagger}_{\alpha\sigma}$($d_{\alpha\sigma}$) are creation (annihilation)
operators in lead-$n$ and the $\alpha$-QD, respectively. Each QD has a 
single particle energy level $\epsilon_{\alpha}$ with spin index $\sigma$,
and the intradot Coulomb interaction is $U_{\alpha}$. To account for 
magnetic field $B$, $\epsilon_{\alpha}$ has a term 
$-\sigma g\mu B_{\alpha}/2$ where $g$ is a constant. We permit 
$U_{upper}\not= U_{lower}$ and $B_{upper}\not=B_{lower}$, 
but these details do not affect our general results. The last term in 
the Hamiltonian describes the coupling between the QDs and the leads, 
and $t_{n,\alpha}$ is the coupling strength. We set 
$t_{1,upper}=t_{3,lower}=0$, meaning there is no coupling between 
the upper-QD and lead-1 and between the lower-QD and lead-3. 

We solve electron current $I_{n,\sigma}$ using standard Keldysh 
nonequilibrium Green's function method (NEGF)\cite{ref5} 
($\hbar=1$):
$
I_{n\sigma}= -2e Im \sum\limits_{\alpha}\int \frac{d\epsilon}{2\pi}
\Gamma_{n\alpha} \left\{ f_n(\epsilon) G^r_{\alpha\sigma}(\epsilon) +
\frac{1}{2} G^<_{\alpha\sigma}(\epsilon)\right\}
$
where $\Gamma_{n,\alpha} \equiv 2\pi \sum_k |t_{n,\alpha}|^2
\delta(\epsilon-\epsilon_{nk})$ is the linewidth function.  
$f_n(\epsilon)$ is the Fermi 
distribution function in lead-$n$. The NEGF 
$G^{r,<}_{\alpha\sigma}(\epsilon)$ is the 
Fourier transform of $G^{r,<}_{\alpha\sigma}(t)$: with 
$G^{r}_{\alpha\sigma}(t) \equiv -i \theta(t) <\{ d_{\alpha\sigma}(t),
d^{\dagger}_{\alpha\sigma}(0)\}>$ and
$G^{<}_{\alpha\sigma}(t) \equiv i  < d^{\dagger}_{\alpha\sigma}(0)
d_{\alpha\sigma}(t)>$.

We solve the retarded Green's function $G^r_{\alpha\sigma}$ in
by the standard equation of motion technique where
indirect tunneling processes such as
upper-QD$\rightarrow$lead-2$\rightarrow$lower-QD are neglected, this is 
reasonable because the long middle barrier between the QDs
helps to block such events to a large extent. We obtain\cite{ref4}:
\begin{equation}
G^r_{\alpha\sigma}(\epsilon) =
 \frac{\epsilon^-_{\alpha\sigma}
     +U_{\alpha}n_{\alpha\bar{\sigma}} }
  {(\epsilon-\epsilon_{\alpha\sigma})\epsilon^-_{\alpha\sigma}
 +\frac{i}{2} \Gamma_{\alpha}
   (\epsilon^-_{\alpha\sigma}+U_{\alpha}n_{\alpha\bar{\sigma}} ) },
\end{equation}
where $\epsilon^-_{\alpha\sigma}\equiv \epsilon-
\epsilon_{\alpha\sigma}-U_{\alpha}$,
$\epsilon_{\alpha\sigma}\equiv\epsilon_{\alpha}-\sigma g\mu 
B_{\alpha}/2$,
$\Gamma_{\alpha} =\sum_n \Gamma_{n\alpha}$,
and $n_{\alpha\bar{\sigma}}$ is the intradot occupation number of
state $\bar{\sigma}$ in the ${\alpha}$-QD. $n_{\alpha\bar{\sigma}}$ 
needs to be calculated self-consistently from the self-consistent 
equation $n_{\alpha\bar{\sigma}} =-i\int \frac{d\epsilon}{2\pi}
G^<_{\alpha\sigma}(\epsilon)$. As usual, $G^r_{n\sigma}(\epsilon)$ has 
two resonances: one at energy $\epsilon_{\alpha\sigma}$ for which the 
associated state $\epsilon_{\alpha\bar{\sigma}}$ is empty; the other is at 
$\epsilon_{\alpha\sigma}+U_{\alpha}$ for which the associated state 
$\epsilon_{\alpha\bar{\sigma}}$ is occupied.

Following the approach of Ref.\onlinecite{ref9}, we obtain
$\int d\epsilon G^<_{\alpha\sigma}(\epsilon)$ 
which is needed in computing current and occupation number:
$$
\int\frac{d\epsilon}{2\pi} G^<_{\alpha\sigma}(\epsilon) =
-\int \frac{d\epsilon}{2\pi}
  \sum\limits_n \frac{\Gamma_{n\alpha}f_n }{\Gamma_{\alpha} }
   (G^r_{\alpha\sigma}(\epsilon)-G^a_{\alpha\sigma}(\epsilon)) \ .
$$
This completes the analytical derivation.

We set bias voltages $\mu_1=0.05$, $\mu_2 =0$, $\mu_3 =-0.05$
so that $\mu_1 > \mu_2 >\mu_3$. We set gate voltages $V_{g,\alpha}$ 
such that at zero magnetic field, $\epsilon_{lower} =\mu_1$
and $\epsilon_{upper}+U_{upper} =\mu_3$. With
this condition there is one electron in the upper-QD.
Fig.(3a,b) shows electron current $I_{n\uparrow}$ and
$I_{n\downarrow}$; charge-current in lead-2 
$I_{2e}=e(I_{2\uparrow}+I_{2\downarrow})$; and 
spin-current in lead-2 
$I_{2s}=\frac{\hbar}{2}(I_{2\uparrow}-I_{2\downarrow})$,
versus a uniform field strength $B$. At zero $B$, 
electron current is non-polarized so that 
$I_{n\uparrow}=I_{n\downarrow}$, and both 
$I_{2e}$ and $I_{2s}$ vanish.
When $B$ increases from zero, the intradot level 
$\epsilon_{\alpha}$ is split. Then levels $\epsilon_{lower,\uparrow}$ 
and $\epsilon_{upper,\downarrow}+U_{upper}$ are moved into the 
bias ``window'' between $\mu_1$($\mu_3$)  and $\mu_2$, while levels 
$\epsilon_{lower,\downarrow}$ and 
$\epsilon_{upper,\uparrow}+U_{upper}$ are moved out of the window,
see Fig.(2). In this situation the electron current in lead-1 and lead-3
are polarized with $I_{\alpha\uparrow}\not= I_{\alpha\downarrow}$. Moreover,
we have $|I_{1\uparrow}| >|I_{1\downarrow}|$ and
$|I_{3\uparrow}| <|I_{3\downarrow}|$. In the following, we focus on
current in lead-2, shown in Fig.(3b). In lead-2 the
value of electron current $I_{2\uparrow}$ equals to the value
of $I_{2\downarrow}$, but their flow direction is exactly 
opposite to each other hence we have $I_{2\uparrow} = 
-I_{2\downarrow}$. 
We therefore obtain zero charge-current $I_{2e}=0$; and a net 
spin-current $I_{2s}$ emerges. When parameter $g\mu B/2 \approx 0.03$, 
the indradot levels $\epsilon_{lower,\uparrow}$ and 
$\epsilon_{upper,\downarrow}+U_{upper}$ are in the middle of the 
bias window, leading to the maximum spin-current.
If field $B$ increases further, the spin-current slightly 
decreases.
 
The device discussed here should be realizable using present 
technology because lateral double-QD structures have 
already been fabricated\cite{andy}. Our analysis
also show that the device does not have a very strict parameter 
requirement. (i) The sizes of the two QDs need not be the same; 
the intradot Coulomb interaction parameters $U_{upper}, U_{lower}$ need 
not be the same. (ii) The field $B$ may or may not be 
uniform, it may also point to any direction. For different directions 
of ${\bf B}$, a spin-current is still induced but the spin polarization 
would depends on the field direction. (iii) The four coupling strengths 
($\Gamma_{1,lower},\Gamma_{2,lower},\Gamma_{3,upper},\Gamma_{2,upper}$) 
between the QDs and the leads can be controlled by split 
gate voltages ($V_{sp1},V_{sp2},V_{sp3},V_{sp4})$ as shown in Fig.(1), 
and they do not need to be the same. In fact, one may fix any three of 
the four and only regulate the last one to obtain a pure 
spin-current with zero charge-current. For example, fixing 
$\Gamma_{1,lower}\neq \Gamma_{2,lower}\neq \Gamma_{2,upper}$, 
the spin-current $I_{2s} $ and charge-current $I_{2e}$ 
versus $\Gamma_{3,upper}$ is shown in Fig.(3c). At a special value of 
$\Gamma_{3,upper}$ given by relation $\Gamma_{3,upper}\Gamma_{2,upper}/
\Gamma_{upper} =\Gamma_{1,lower}\Gamma_{2,lower} /\Gamma_{lower}$, 
$I_{2e}$ vanishes and only $I_{2s}$ exists. (iv) So far we have set 
$\epsilon_{lower} =\mu_1$ and $\epsilon_{upper}+U_{upper} =\mu_3$, but 
these conditions can be relaxed. For example, if 
$\epsilon_{upper}+U_{upper}=-0.06$, 
somewhat different from $\mu_3$, by regulating the lower-QD level 
$\epsilon_{lower}$ using gate voltage $V_{g,lower}$, we can easily find 
the operation point for large $I_{2s}$ with zero $I_{2e}$, as shown in 
Fig.(3d). (v) As for the parameter values, in Fig.(3) we have used 
$k_B T =0.01$. Assuming this is equivalent to 100mK,\cite{addnote1}
other parameter 
values used to generate Fig.(3) can be deduced. We find: 
$V_1 =\mu_1 /e \approx 43 \mu V$, $V_2 =0$,
$V_3 \approx -43 \mu V$, $U_{\alpha} \approx 1 meV$, and 
$B\approx 0.8/g$ tesla for $g\mu B/2 =0.03$.\cite{addnote2}
These parameters are in the
standard range of QD devices.\cite{addnote}

Finally, we discuss in what sense the proposed device behaves as a
spin-battery with two poles. Note that the region indicated by the dotted
box in Fig.(1) is reserved for spintronic devices: any application of
spin-current should be done in this region. The lateral-QD plus the
external circuit constitute the spin-battery: the two poles of the
spin-battery are points ``A" and ``B" as shown in Fig.(1).
If there exists direct connection between A and B, a spin current
is driven through by the spin-battery. On the other hand,
if there is no direct connection, a spin-motive force will be established
between A and B. Importantly, even if there are not spin flip
mechanisms in whole device, the spin-battery can still work, which is
different from the one-pole systems.\cite{ref2,ref3}
Finally, the distance between points A and B can be as large as the
spin coherence length which can reach many $\mu m$ at low
temperatures.\cite{ref11,ref12} Such as large distance should allow
useful applications of the flowing spin-current.

In summary, we have shown that gate-controllable spin-battery for 
spin-current is possible. Such a device should be fabricable using  
present technology. We believe the present design to be superior 
as no time-dependent field is involved. In the present work, we 
did not discuss detection of pure spin-current without charge-current, 
but such discussions already exist in literature\cite{heinrich,ref11,ref13} 
and we refer interested readers to them.

{\bf Acknowledgments:} 
We gratefully acknowledge financial support from NSERC of Canada, FCAR 
of Quebec (Q.S., H.G), and a RGC grant from the SAR Government of Hong 
Kong under grant number HKU 7091/01P (J.W.). 


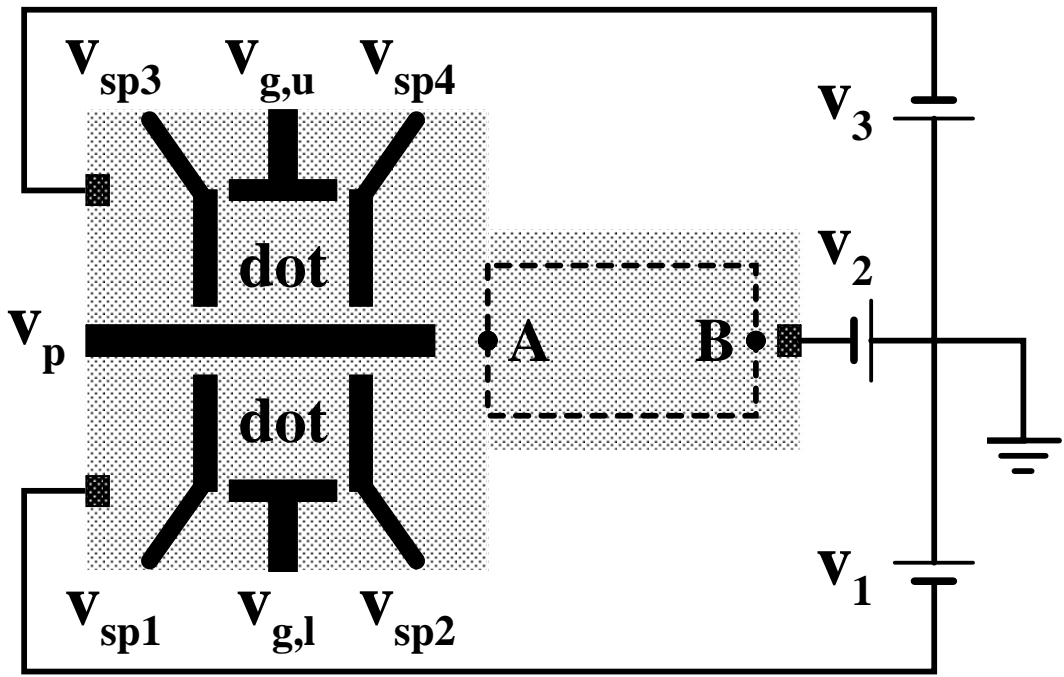
\begin{figure}
\caption{
Schematic diagram for the lateral quantum-dot. The lightly shaded
region represents two-dimensional electron gas, the darker regions are
the metal gates (including split gates $V_{spn}$, $V_p$, and 
gate voltage $V_{g,\alpha}$). The dotted box represents the region in
which a pure spin-current flows through.
}
\label{fig1}
\end{figure}

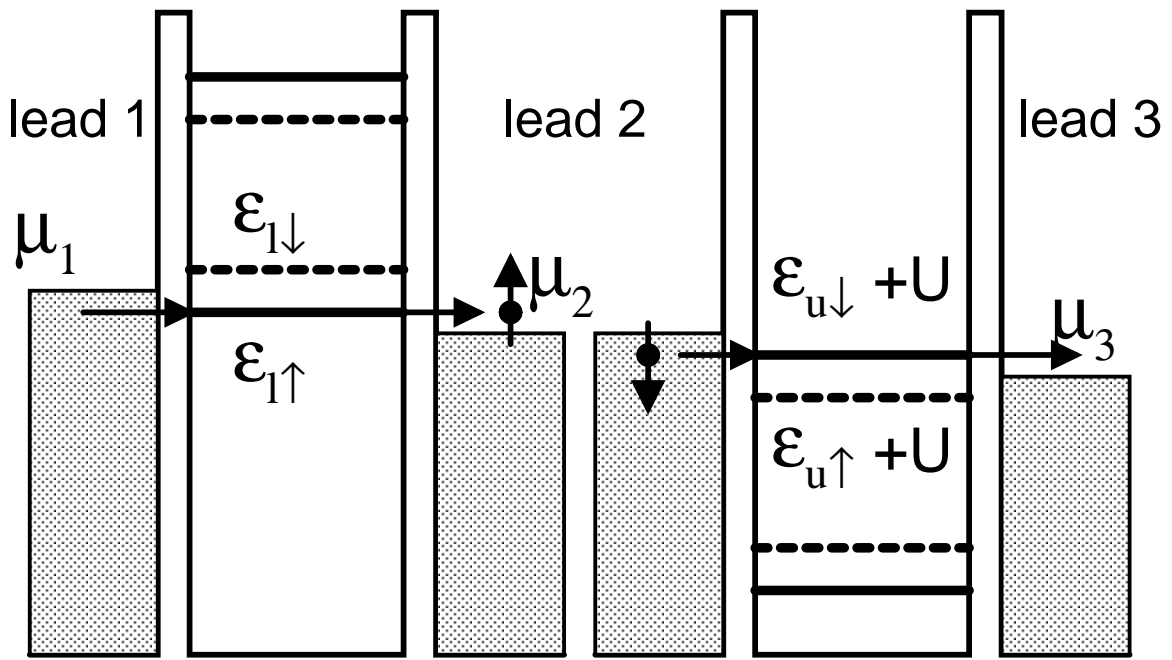
\begin{figure} 
\caption{ 
Schematic plot of energy level position and the tunneling process
during spin-battery operation.
} 
\label{fig2}
\end{figure}

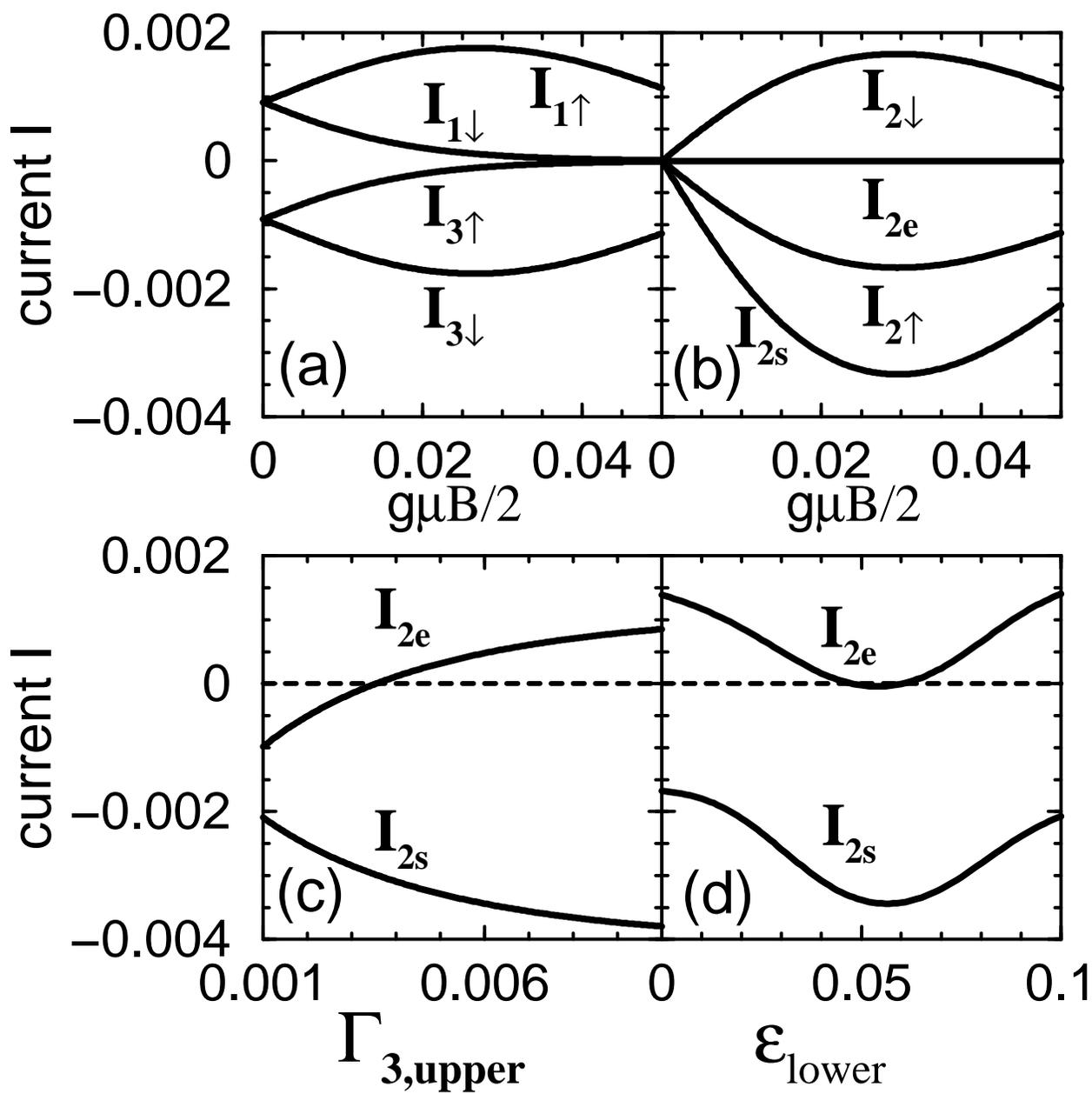
\begin{figure} 
\caption{ 
(a) and (b): for electron currents $I_{n\uparrow}$ and 
$I_{n\downarrow}$,
charge-current $I_{2e}$ (unit $e$), and spin-current $I_{2s}$ 
(unit $\frac{\hbar}{2}$), versus magnetic field parameter $g\mu B/2$.
Other parameters are 
$\Gamma_{1,lower} =\Gamma_{2,lower} =\Gamma_{2,upper} =
\Gamma_{3,upper} =0.005$, $k_B T=0.01$, $U_{lower}=1.0$, 
$U_{upper}=0.9$, 
$\epsilon_{lower}=\mu_1$, and $\epsilon_{upper}+U_{upper} =\mu_3$.
(c): $I_{2e}$ and $I_{2s}$ versus $\Gamma_{3,upper}$ 
with $g\mu B/2 =0.03$. 
Other parameters are: $\Gamma_{1,lower} =0.004$, 
$\Gamma_{2,lower} =0.005$, 
and $\Gamma_{2,upper} =0.006$. (d): $I_{2e}$ and $I_{2s}$ versus 
$\epsilon_{lower}$ with $g\mu B/2 =0.03$. Here 
$\epsilon_{upper}+U_{upper} 
=-0.06$ which is slightly different from $\mu_3 =-0.05$.
Other parameters in (c) and (d) are the same as those in (a) and (b).
} 
\label{fig3}
\end{figure}


\begin{references}
\bibitem{addnote}
In our model Homiltonian, we consider only one level for each QD. 
This is reasonable if the level spacing $\Delta$ is larger than the bias 
$|V_{1/3} -V_2|$. This should be realizable in present technology.
For example, for a $0.2\mu m \times 0.2 \mu m$ QD, its $\Delta$
is about $\pi \hbar /(m^* area) \approx 90 \mu V > |V_{1/3}- V_2| 
\approx 43 \mu V$.
\end{references}
\end{document}